\documentstyle[prl,aps,preprint,floats,amssymb,graphicx]{revtex}

\newcommand{\pvalt}{\raise0.15ex\hbox{-}\mkern-11.5mu\int}
\newcommand{\be}{\begin{equation}}
\newcommand{\ee}{\end{equation}}
\newcommand{\bea}{\begin{eqnarray}}
\newcommand{\eea}{\end{eqnarray}}
\newcommand{\ben}{\begin{enumerate}}
\newcommand{\een}{\end{enumerate}}
\newcommand{\bit}{\begin{itemize}}
\newcommand{\eit}{\end{itemize}}
\newcommand{\0}{0^{++}}
\newcommand{\la}[1]{\label{#1}}

\newcommand{\eq}[1]{eq.~(\ref{#1})}

\newcommand{\half}{\frac{1}{2}}

\newcommand{\Tr}{\,\mbox{Tr}\,}

\newcommand\qq{\bar q q}
\newcommand\qqqq{\bar q^{2}q^{2}}

\newcommand{\ga}{\gamma}

\newcommand{\de}{\delta}

\newcommand{\ka}{\kappa}

\renewcommand{\O}{{\cal O}}
\newcommand{\fm}{{\rm fm}}
\newcommand{\GeV}{{\rm GeV}}
\newcommand{\MeV}{{\rm MeV}}

\newcommand{\beq}{\begin{equation}}
\newcommand{\eeq}{\end{equation}}
\newcommand{\ba}{\begin{array}}
\newcommand{\ea}{\end{array}}
\newcommand{\dsp}{\displaystyle}
 
\renewcommand{\>}{\rangle} 
\newcommand{\exotic}{{\rm E}}
\newcommand{\nonexotic}{{\rm N}}
\newcommand{\dz}{\de E_\nonexotic}
\newcommand{\dt}{\de E_\exotic}
\newcommand{\etal}{{et~al.}}

\begin{document}
\tighten

\setcounter{page}{0}


\preprint{\vbox{MIT-CTP-2940
                        \null\hfill\rm \today} }

\title{Insight into the Scalar Mesons from a Lattice Calculation}

\author{Mark~Alford and R.~L.~Jaffe}

\address{{~}\\
Center for Theoretical Physics,\\
   Laboratory for Nuclear Science
   and Department of Physics,\\
   Massachusetts Institute of Technology,\\
   Cambridge, Massachusetts 02139\\
{~}}

\maketitle

\begin{abstract}

We study the possibility that the light scalar mesons are $\qqqq$
states rather than $\qq$. We perform a lattice QCD calculation
of pseudoscalar meson scattering amplitudes,
ignoring quark loops and quark annihilation, and find indications that
for sufficiently heavy quarks there is a stable four-quark bound state
with $J^{PC}=0^{++}$ and non-exotic flavor quantum numbers.

\end{abstract}

\narrowtext
\newpage

\section{Introduction}
\label{section0}

The light scalar mesons have defied classification for
decades \cite{Caso:1998,Black:1999}. Some are narrow and have been
firmly established since the 1960's.  Others are so broad that their
very existence is controversial.  Scalar mesons are predicted to be
chiral partners of the pseudoscalars like the pion, but their role in
chiral dynamics remains obscure.  Naive quark models interpret them as
orbitally excited $\qq$ states.  Others have suggested that they are
$\qqqq$\cite{Jaffe:1977} or ``molecular'' states,\cite{Weinstein:1990}
strongly coupled to $\pi\pi$ and $\bar K K$ thresholds.

In this paper we propose a 
way to shed some light on the nature of the scalar mesons 
using lattice QCD. Previously scalar mesons have been treated like 
other mesons: their masses have been 
extracted from the large Euclidean time falloff of 
$\qq-\qq $ correlation functions with the 
appropriate quantum numbers.  Here we look for a $\0\,\qqqq$ 
\emph{bound state}. We construct $\qqqq$ sources, work 
in the quenched approximation,
and discard $\qq$ annihilation diagrams
so communication with $\qq$ and vacuum 
channels is forbidden.  Also, we allow the quark masses to be large 
(hundreds of MeV), so the continuum threshold for the decay $\qqqq\to 
(\qq) (\qq)$ is artificially elevated.  We then study the 
large Euclidean time falloff of a $\qqqq-\qqqq$ correlator, looking 
for a falloff slower than $2m_{\qq}$, signalling a bound 
state. Such an object would have been missed by studies 
of $\qq$ correlators in the quenched approximation.  We use 
shortcomings of lattice QCD to our advantage. By excluding
processes that mix $\qq$ and $\qqqq$, we can
unambiguously assign a quark content to a state.  The heavy 
quark mass suppresses relativistic effects, which we believe complicate the 
interpretation of light quark states.

Our initial results are encouraging: within the limits of our
computation we see signs of a bound state in the ``non-exotic'' $\qqqq$
channel, namely, the one with quantum numbers that could also
characterize a $\qq$ state ($I=0$ for 2 flavors,
the {\bf 1} and {\bf 8} for 3 flavors).  In contrast,
the ``exotic'' flavor $\qqqq$ channel ($I=2$ for 2 flavors,
the {\bf 27}  for 3 flavors) shows 
no bound state.  Instead it shows a negative scattering
length, characteristic of a repulsive interaction.  To obtain a
definitive result will require larger lattices and more computer time,
but this is well within the scope of existing facilities.

In Sec.~\ref{section1} we give an overview of the $\0$ mesons.  First
we summarize the phenomenology.  Then we summarize previous lattice
calculations.  We also review earlier studies of $\qqqq$ sources on
the lattice.\cite{GPS2,Fukugita} Because these earlier works looked only at
one (relatively small) lattice size they were unable to examine the
possibility of a bound state.  In Sec.~\ref{section2} we summarize our
computation.  First we briefly review $\qqqq$ operators and discuss
lattice size and quark mass dependence.  Next, we review the improved
lattice action we use to enable us to study larger lattices.\cite{land234}
Finally in Sec.~\ref{section3} we present our results and discuss
their implications.  We explore some of the directions in which our 
computation could be improved.

A reader who wishes to skip the details can look immediately at 
Fig.~\ref{fig:falloff} where we 
plot the dependence on lattice size of the binding energy associated 
with the exotic and non-exotic $\qqqq$ channels.  The exotic channel 
shows a negative binding energy with the $1/L^{3}$ dependence expected 
from analysis of the $(\qq)(\qq)$ continuum.\cite{Luscher} The 
coefficient of $1/L^{3}$ agrees 
roughly with Refs.~\cite{GPS2,Fukugita} and 
with the predictions of chiral perturbation theory.  The non-exotic 
channel shows positive binding energy, but seems to depart from 
$1/L^{3}$, perhaps approaching a constant as $L\to\infty$, which 
would indicate the existence of a bound 
$\qqqq$ state.  Confirmation of this result will require 
further calculations on larger lattices.

\section{Overview of the light scalar mesons}
\label{section1}

In this section we establish the context for our work.  First we give
a very brief introduction to the phenomenology of the lightest $\0$
mesons composed of light ($u$, $d$, and $s$) quarks.  We give a sketch
of the  $\qq$ and $\qqqq$ models for $\0$ states and contrast them.  More
information can be found in Refs.~\cite{Caso:1998,Black:1999} and
references quoted therein.  Next we summarize existing lattice
calculations which relate to the $\0$ channel.  These fall into two
classes: traditional searches for $\qq$ eigenstates and attempts to
learn about low energy $\pi\pi$ scattering by studying $\qqqq$
sources.

\subsection{Phenomenology}

The known $\0$ mesons divide into effects near and below 1 GeV, 
which are unusual, and effects in the 1.3--1.5 GeV region which may be 
more conventional.  Here we focus on the states below 1 GeV. 
Altogether, the objects below $1~\GeV$ form an $SU(3)_{\rm f}$ nonet: two 
isosinglets, an isotriplet and two strange isodoublets.  The 
isotriplet and one isosinglet are narrow and well confirmed.  The 
isodoublets and the other isosinglet are very broad and still 
controversial.

The well established $0^{++}$ mesons are the isosinglet $f_{0}(980)$
and the isotriplet $a_{0}(980)$.  Both are relatively narrow:
$\Gamma[f_{0}]\sim$ 40 MeV, $\Gamma[a_{0}]\sim$ 50 MeV,\footnote{We
use the observed peak width into $\pi\pi$ and $\pi\eta$ respectively,
rather than some more model dependent method for extracting a width.}
despite the presence of open channels ($\pi\pi$ for the $f_{0}$ and
$\pi\eta$ for the $a_{0}$) for allowed s-wave decays.   Both couple
strongly to $\bar K K$ and lie so close to the $\bar K K$ threshold
at 987 MeV that their shapes are strongly distorted by threshold
effects.  Interpretation of the $f_{0}$ and $a_{0}$ requires a
coupled channel scattering analysis.  The relevant channels are 
$\pi\pi$ and $\bar K K$ for the $f_{0}$ and $\pi\eta$ and $\bar K K$ 
for the $a_{0}$.  In both cases the results favor an intrinsically 
broad state, strongly coupled to $\bar K K$ and weakly coupled to the 
other channel.  The physical object appears narrow because the $\bar 
K K$ channel is closed over a significant portion of the object's width.
No summary this brief does justice to the wealth of work and 
opinion in this complex situation.

The other light scalar mesons are known as broad enhancements in very
low energy s-wave meson-meson scattering.  The enhancements are
universally accepted, but their interpretation is more controversial. 
At the lowest energies only the $\pi\pi$ channel is open.  The
$\pi\pi$ s-wave can couple either to isospin zero or two.  The $I=2$
(e.g.  $\pi^{+}\pi^{+}$) channel shows a weak repulsion in rough
agreement with the predictions of chiral low energy
theorems.\cite{Weinberg} The $I=0$ channel shows a strong
attraction: the phase shift rises steadily from threshold to
approximately $\pi/2$ by $\sim 800$ MeV before effects associated with
the $f_{0}$ complicate the picture.  This low mass enhancement in the
$\pi\pi$ s-wave is the $\sigma$ meson of nuclear physics and chiral
dynamics.  Recent studies support the existence of an S-matrix pole
associated with this state at a mass around 600 MeV, which we will
refer to as the $\sigma(600)$.\cite{Black:1999,Tornqvist:1996ay} The $\pi
K$ s-wave is very similar to $\pi\pi$.  The exotic $I=3/2$ (e.g. 
$\pi^{+} K^{+}$) channel shows weak repulsion.  The non-exotic $I=1/2$
channel shows relatively strong attraction.  Black~\etal
\cite{Black:1999} identify the
enhancement with an S-matrix pole at approximately 900 MeV, which is
known as the $\kappa$(900).The enhancement is not in doubt, but the
interpretation is, if anything, more controversial than the $\pi\pi$
case.\footnote{For example, the $\kappa$(900) is not mentioned in
Ref.~\cite{Caso:1998}} Other couplings of these objects (the $\sigma$
can couple to $\bar K K$ and the $\kappa$ can couple to $\eta K$) are
unknown because the relevant thresholds lie above the states.  The
large widths of these states reflect their strong coupling to the
open decay channels $\pi\pi$ and $\pi K$ respectively.

The conventional quark model assigns the $\0$ mesons to the first 
orbitally excited multiplet of $\qq$ states.  As in positronium, $\0$ 
quantum numbers are made by coupling $L=1$ to $S=1$ to give total 
$J=0$.  The $\0$ states should be very similar to the $1^{+\pm}$ and 
$2^{++}$ $\qq$ states that lie in the same family.  These are very 
well known and form conventional meson nonets (in $SU(3)_{\rm f}$).  
Since they have a unit of excitation (orbital angular momentum), they 
are expected to be quite a bit heavier than the pseudoscalar and 
vector mesons.  Most models put the $\qq~\0$ mesons along with 
their $2^{++}$ and $1^{++}$ brethren around 1.2--1.5 GeV. 

\begin{figure}
\includegraphics[width=6.5in]{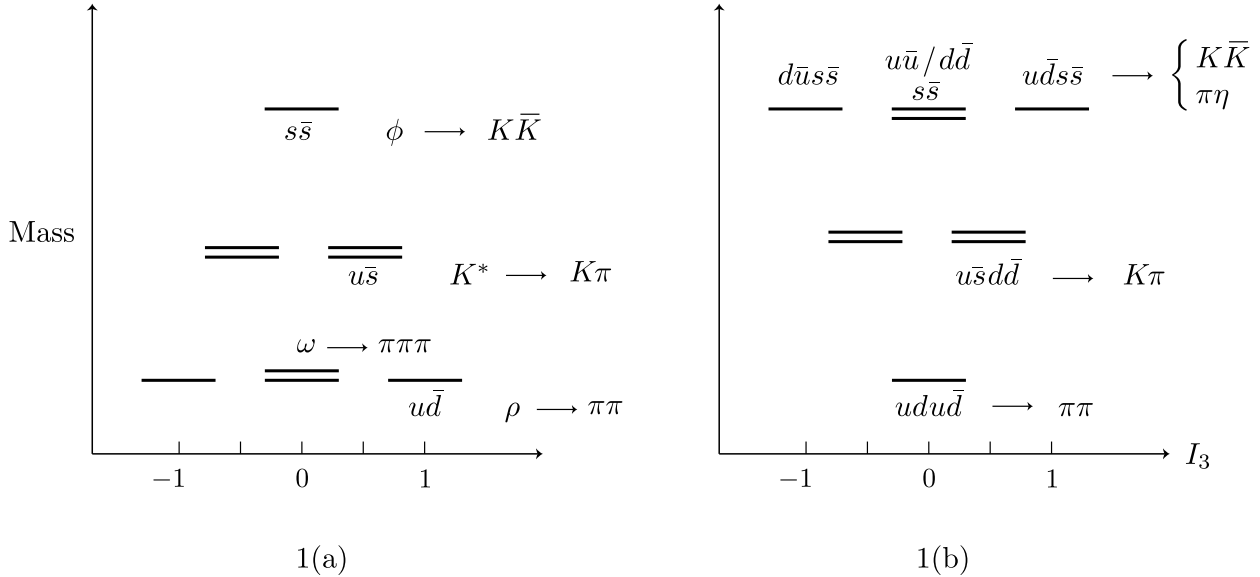}
\vspace{3mm}
\caption{The mass pattern, quark 
content and natural decay couplings of (a) a $\qq$ nonet
and (b) a $\qqqq$ nonet.}
\label{quark}
\end{figure}

An idealized $\qq$ meson nonet has a characteristic pattern of 
masses and decay couplings.  The vector mesons are best known, but the 
pattern is equally apparent in the $2^{++}$ or $1^{++}$ nonets.  The 
isotriplet and the isosinglet composed of non-strange quarks are 
lightest and are roughly degenerate (e.g.~the $\rho$ and the $\omega$).  
The strange isodoublets are heavier because they contain a single 
strange quark (e.g.~the $K^{\ast}$).  The final isosinglet is heaviest 
because it contains an $\bar s s$ pair (e.g.~the $\phi$).  Decay 
patterns show selection rules which follow from this quark content.  
In particular, the lone isosinglet does not couple to non-strange 
mesons ($\phi\not\!\to 3\pi$). 
The mass pattern, quark 
content and natural decay couplings of a $\qq$ nonet are summarized in 
Fig.~\ref{quark}a.  
These patterns seem to bear  little
resemblance to the masses and couplings of the light 
$\0$ mesons, a fact which led earlier workers to explore other
interpretations.

Four quarks ($\qqqq$) can couple to $\0$ without a unit of orbital
excitation.  Furthermore, the color and spin dependent interactions,
which arise from one gluon exchange, favor states in which quarks and
antiquarks are separately antisymmetrized in flavor.  For quarks in
3-flavor QCD the antisymmetric state is the 
flavor $\bar {\bf 3}$.  Thus
the energetically favored configuration for $\qqqq$ in flavor is
$(\bar q\bar q)^{{\bf 3}}(qq)^{\bar {\bf 3}}$,
a flavor nonet. 
The lightest multiplet has spin 0.  Explicit studies in the MIT Bag
Model indicated that the color-spin interaction could drive the
$\qqqq$ $\0$ nonet down to very low energies: 600 to 1000 MeV
depending on the strangeness content.\cite{Jaffe:1977}

The most striking feature of a $\qqqq$ nonet in comparison with a
$\qq$ nonet is an \emph{inverted mass spectrum}
(see Fig.~\ref{quark}b).  The crucial
ingredient is the presence of a hidden $\bar s s$ pair in several
states.  The flavor content of $(qq)^{\bar {\bf 3}}$ is $\{[ud],
[us], [ds]\}$, where the brackets denote antisymmetry.  When combined
with $(\bar q\bar q)^{{\bf 3}}$, four of the resulting states
contain a hidden $\bar s s$ pair: the isotriplet and one of the
isosinglets have quark content $\{u\bar d s \bar s,
\frac{1}{\sqrt{2}}(u\bar u-d\bar d)s\bar s, d\bar u s \bar s\}$ and
$\frac{1}{\sqrt{2}}(u\bar u+d\bar d)s\bar s$, and therefore lie at the
top of the multiplet.  The other isosinglet, $u\bar d d \bar u$ is the
only state without strange quarks and therefore lies alone at the
bottom of the multiplet.  The strange isodoublets ($u\bar s d\bar d$,
etc.)  should lie in between.  In summary, one expects a degenerate
isosinglet and isotriplet at the top of the multiplet and strongly
coupled to $\bar K K$, an isosinglet at the bottom, strongly coupled to
$\pi\pi$, and a strange isodoublet coupled to $K\pi$ in between 
(Fig.~\ref{quark}b).  The resemblance to the observed structure of the 
light $\0$ states is considerable.  

These qualitative considerations motivate a careful look at the 
classification of the scalar mesons.  Models of QCD are not 
sophisticated enough to settle the question.  For example, the $\qqqq$ 
picture does not distinguish between one extreme where the four quarks 
sit in the lowest orbital of some mean field,\cite{Jaffe:1977} and the 
other, where the four quarks are correlated into two $\qq$ mesons 
which attract one another in the flavor $(\bar q\bar q)^{{\bf 3}} 
(qq)^{\bar {\bf 3}}$ channel.\cite{Jaffe:1979bu,Weinstein:1990} For 
years, phenomenologists have attempted to analyse meson-meson 
scattering data in ways which might distinguish between $\qq$ and 
$\qqqq$ assignments.  A recent quantitative study favors the $\qqqq$ 
assignment.\cite{Black:1999} However the $\qq$ assignment has strong 
advocates.\cite{Tornqvist:1996ay} We hope that a suitably constrained 
lattice calculation can aid in the eventual classification of these 
states.

\subsection{Existing Lattice Studies}

In this section we briefly summarize existing lattice calculations 
which bear on the classification of the $\0$ mesons. 
There have been lattice studies of both the spectrum of $\0$
states and the mixing of $\qq$ states with glueballs.

Unquenched spectroscopic calculations are just beginning to become
available\cite{SESAM,UKQCD2}. 
In principle, they are of interest because they would
couple to a $\qqqq$ configuration if it is energetically
favorable. One unquenched calculation reports tentative evidence of
$\0$ state at an energy much lower than that reported in quenched
calculations \cite{UKQCD2}. We return to this work briefly in our
conclusions. Further insight from unquenched calculations will have
to await more definitive studies.

For the rest of this section, we restrict ourselves to consideration
of quenched calculations.  We will not discuss the mixing of glueballs
with $\qq$ states, because we are interested in distinguishing $\qq$
from $\qqqq$ components of mesons.  First we consider quenched studies
of $\qq$ spectra.  Then we describe attempts to extract meson-meson
scattering lengths from quenched studies of $\qqqq$ sources.

\subsubsection{$\qq$ Spectrum Calculations}

\begin{figure}
\includegraphics[width=3.5in,angle=-90]{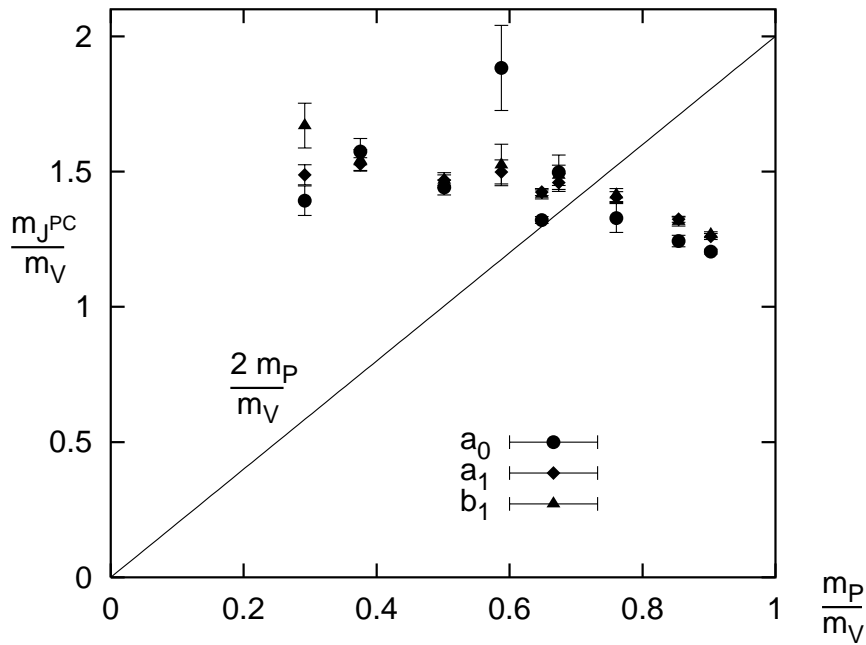}
\vspace{3ex}
\caption{ Quenched lattice calculations \protect\cite{KimOhta,qcdsf}
of the ratio of the masses of various $p$-wave $\qq$ mesons to the vector
meson mass, as a function of the pseudoscalar to vector mass
ratio, ie with varying quark mass. The threshold for decay to
$P\,P$ is also shown.
}
\label{fig:qqbar}
\end{figure}

The masses of the $\0$ $\qq$ states have been calculated on the
lattice in the quenched approximation by various groups
\cite{LeeWein,KimOhta,UKQCD1,qcdsf}.
Some of their results are shown in Fig.~\ref{fig:qqbar}.  
As well as the $J^{PC}=\0 (a_0)$,
we have included data from the same groups on the other
positive parity mesons  $J^{PC}=1^{++}(a_{1})$ and $1^{+-}(b_{1})$. 
Data for the $2^{++}(a_{2})$ was not available.  
The spectra in Fig.~\ref{fig:qqbar} 
behave roughly as $\qq$ states with orbital angular momentum should.  In the 
heavy quark limit, as 
the pseudoscalar mass $m_P$ approaches the vector mass $m_V$, their masses 
approach one another because spin and spin-orbit splittings decrease with 
$m_{q}$, and approach $m_V$ because orbital excitation energy 
decreases with $m_{q}$.  To make this behavior manifest we plot 
$m_{J^{PC}}/m_V$ versus $m_P/m_V$, and
note that $m_{J^{PC}}/m_V$  approaches unity as $m_P/m_V$ increases.

\subsubsection{Pseudoscalar Scattering Length Calculations}

In the past, lattice studies of four-quark states have been undertaken
in order to extract pseudoscalar-pseudoscalar ($P$-$P$) scattering
lengths for comparison with the predictions of chiral dynamics. It is
known \cite{Luscher} that the energy shift $\de E$ of a two-particle
state with quantum numbers $\alpha$ in a cubic box of size $L$ is
related to the threshold scattering amplitude,
\beq 
\label{luscher} 
\de E_{\alpha} = E_{\alpha} - 2 m_P = 
\frac{T_{\alpha}}{L^3}\left(  1 + 2.8373\, \frac{m_P T_{\alpha}}{4 \pi L}
+ 6.3752 \left(\frac{m_P T_{\alpha}}{4 \pi L}\right)^{\!2}
 + \cdots \right), 
\eeq
where $m_P$ is the mass of the scattering particles, and $T_{\alpha}$ is 
the scattering amplitude at threshold
in the channel labelled by $\alpha$, which can be related to
the scattering length, 
\be
T_{\alpha}=-\frac{4\pi a_{\alpha}}{m_P}.
\la{scatt-length}
\ee
For a more detailed discussion, see Ref.~\cite{GPS1}. In our case the 
channels of interest are exotic ($I=2$, for two flavors) and non-exotic 
($I=0$, for two flavors).  If the interaction is attractive enough to 
produce a bound state, then instead of \eq{luscher} one would find that 
$\de E$ goes to a negative constant as $L\to \infty$.

In order to distinguish between a bound state and the continuum 
behavior described by \eq{luscher}, it is necessary to perform 
calculations for several different lattice sizes.  Calculations with 
$\qqqq$ sources have been performed by Gupta~\etal \cite{GPS2}, who 
studied one lattice volume at one lattice spacing, and Fukugita~\etal 
\cite{Fukugita}, who, for the heavy quark masses we are interested in, 
also studied only one lattice volume at one lattice spacing.  Their 
results were therefore not sufficient to check the lattice-size 
dependence of energy of the two-pseudoscalar state, 
and investigate the possibility of a bound state.  
Our method follows theirs, but we have studied a range of lattice sizes.
Their results are plotted along with ours in 
Fig.~\ref{fig:falloff}.  Where our calculations overlap, they agree.

\section{A ${\mathbf \qqqq}$ Exercise on the Lattice}
\label{section2}

\subsection{Quark contractions and flavor dependence}
\label{sec:contractions}

\begin{figure}
\includegraphics[width=5.5in]{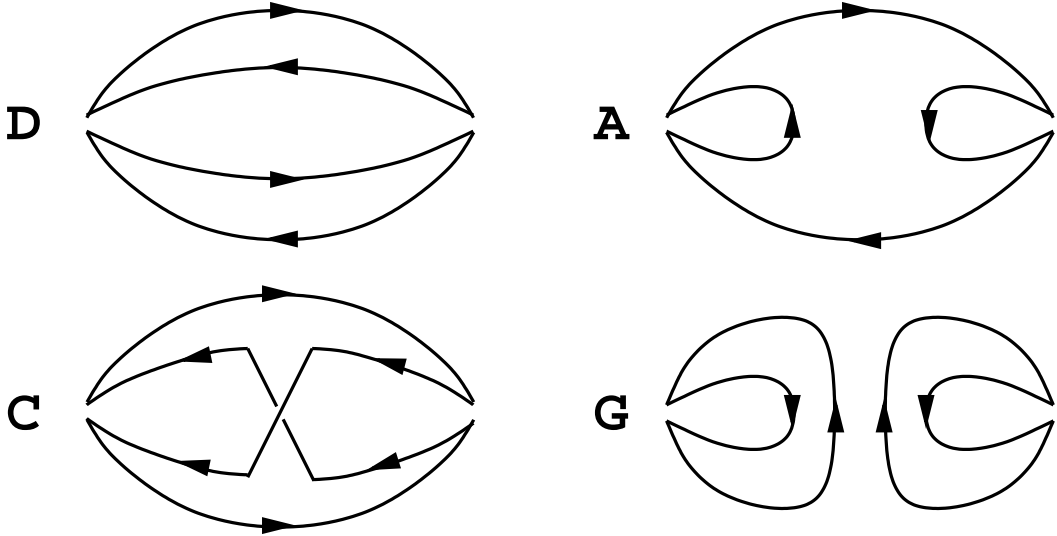}
\vspace{3ex}
\caption{ The four types of quark line contraction that contribute to
the pseudoscalar-pseudoscalar ($P$-$P$) correlation function.
}
\label{fig:dcag}
\end{figure}

For our purposes the salient categorization of
$\qqqq$ correlators is into ``exotic''
channels (flavor states that are only possible for a $\qqqq$ state,
$I=2$ for two flavors, the {\bf 27} for three flavors) and 
non-exotic channels (flavor states that could be $\qqqq$ or $\qq$,
$I=0$ for two flavors, the {\bf 8} and {\bf 1} for three flavors).
In the absence of quark annihilation diagrams, the {\bf 8} and {\bf 1}
are identical. When annihilation is included, the  {\bf 1}, like the
$I=0$ for two flavors, can mix with pure glue.
As shown in Fig.~\ref{fig:dcag}, the $\qqqq~\0$ correlation functions can be
expressed in terms of a basis determined by the
four ways of contracting the quark
propagators\cite{GPS1}: direct (D), crossed (C), single annihilation (A),
complete annihilation into glue (G).
Since we are interested in $\qqqq$ states, we only study the D and C
contributions.  We will assume that all quarks are degenerate, so
there is only one quark mass, and as far as color and spinor indices
are concerned all quark propagators are the same. In our lattice
calculation we will therefore build our $\qqqq$ correlators from color
and spinor traces of contractions of four identical quark
propagators, putting in the flavor properties by hand when we choose
the relative weights of the different contractions.

In the case of two flavors, there are two possible channels
for a spatially symmetric source: $I=2$ (exotic) and $I=0$ (non-exotic).
Evaluation of the flavor dependence of the quark line contractions shows that
the $I=2$ channel is $D-C$, and $I=0$ is $D+\half C$ \cite{GPS1}.

For three flavors, the possible channels are the symmetric
parts of ${\bf 3}\times{\bf 3}\times\bar{\bf 3}\times\bar{\bf 3}$,
namely ${\bf 1}+{\bf 8}$ (non-exotic) and ${\bf 27}$ (exotic).
As in the two-flavor case, the exotic channel is $D-C$.
At sufficiently large Euclidean time separation, each
contraction will behave as a sum of exponentials, corresponding
to the states it overlaps with. Generically, all linear combinations
will be dominated by the same state: the lightest. Only with
correctly chosen relative weightings
will the leading exponential cancel out, yielding a faster-dropping
exponential corresponding to a more massive state.
We will see in Sec.~\ref{sec:results} that the exotic
($D-C$) channel is the one where such a cancellation occurs, yielding
a repulsive interaction between the pseudoscalars.
For {\em any} other linear combination of $D$ and $C$ the
correlator is therefore dominated by the lightest, attractive state.
Without loss of generality, we can therefore study the following
linear combinations:
\beq
\ba{r@{\qquad}rcl@{\qquad\mbox{2 flavor:~}}l@{\qquad\mbox{3 flavor:~}}l } 
\mbox{Exotic:} & J_\exotic &=& D - C           & I=2 & {\bf 27} \\
\mbox{Non-exotic:} & J_\nonexotic &=& D + \half C  & I=0 & {\bf 1}, {\bf 8}
\ea
\eeq

We conclude that if, as our results suggest, there is a bound $\qqqq$ state
in the non-exotic channel, then this means that with two flavors,
the $I=0$ channel is bound, and with three flavors both the
{\bf 1} and {\bf 8} are bound.
Once quark loops and annihilation diagrams are included, the
{\bf 1} and {\bf 8} will split apart. Unquenched lattice calculations
will be needed to see if they remain bound.

\subsection{Lattice action}

In our lattice calculations, we work in the quenched (valence)
approximation, and use Symanzik-improved glue and quark actions. This
means that irrelevant terms ($\O(a), \O(a^2)$, where $a$ is the
lattice spacing) have been added to the lattice action to compensate
for discretization errors.

Improved actions are crucial to our ability to explore a range
of physical volumes using limited computer resources.
Because most of the finite-lattice-spacing errors have been
removed, we can use coarse lattices, which have fewer sites
and hence require much less computational effort: note that
the number of floating-point-operations  required
even for a quenched lattice QCD calculation rises faster than
$a^{-4}$.

Improved actions have been studied extensively
\cite{land234,WW,LW,Alf1,Improving}, and it has been found that
even on fairly coarse lattices ($a$ up to $0.4~\fm$)
good results can be obtained for hadron masses by estimating
the coefficients of the improvement terms using tadpole-improved
perturbation theory.
For the energy differences 
that we measure, we find that the improved
action works very well. There are
no signs of lattice-spacing dependence at $a$ up to $0.4~\fm$, so
as well as greatly reducing the computer resources required, it
enables us to dispense with the extrapolation in $a$
that is usually needed to obtain continuum results.

Our lattice glue and quark action parameters are summarized in table
\ref{tab:params} and are described in detail in Ref.~\cite{land234}.
For the glue we use a L\"uscher-Weisz (plaquette and $2\!\times\!1$
rectangle) action \cite{LW,Alf1}.  We measured the lattice spacing by NRQCD
calculations of the charmonium $P\!-\!S$ splitting, using the experimental
spin-averaged value of $458~\MeV$.

For the quarks we use a D234 action, which includes third and fourth
derivative terms as well as an improved clover term.  All the
coefficients in the action are evaluated at tree level in
tadpole-improved perturbation theory \cite{LM}, using the mean link in Landau
gauge to estimate the tadpole contribution.  We work at a quark mass
close to the physical strange quark: the pseudoscalar to vector meson
mass ratio $m_P/m_V$ is 0.76.

We collected data at two lattice spacings, $a=0.40$ and $0.25~\fm$.
The scaling of the hadron masses is good but not perfect: the pseudoscalar
weighed $790~\MeV$ on our coarser lattice and $840~\MeV$ on the
finer one. For our fits to \eq{luscher}
in Sect.~\ref{sec:results} we used the average.

\subsection{Sources and fitting}
\label{sources}

To look for bound $\0~\qqqq$ states, we investigate
states of two pseudoscalar mesons on lattices of various volumes,
keeping only quark-line-connected diagrams.  We 
calculate the binding energy $\dt$ in the exotic
channel (flavor states that are only possible for a $\qqqq$ state, ie
$I=2$ for two flavors, the {\bf 27} for three flavors) and 
the binding energy $\dz$ in the
non-exotic channel (flavor states that could be $\qqqq$ or $\qq$,
ie $I=0$ for two flavors, the {\bf 8} and {\bf 1} for three flavors).
We could have used sources based on preconceptions about
maximally attractive channels in QCD. For example, one-gluon exchange
and instanton interactions are known to favor the color $\bar {\bf 3}$
diquark channel, leading to interesting phenomenology at high
density\cite{colorSC}, but we verified that these have good overlap with
the two pseudoscalar meson source that we used.

Since we are interested in $\qqqq$ states, we only study the D and C
contributions. 
For each gauge field configuration we evaluate
the pseudoscalar correlator $P(t)$ and the
the direct (``$D$'') and crossed (``$C$'') contributions 
to the two-pseudoscalar correlator.  We use a
wall source at $t=0$, and both point and smeared sinks.
\newcommand{\GW}{G}
\beq\label{contractions}
\ba{rcl}
P(t) &=& \dsp \sum_{\vec x\/} \Tr\Bigl( 
  \GW(t,\vec x)\, \GW^\dagger(t,\vec x\/) \Bigr) \\
D(t) &=& \dsp \sum_{\vec x\/} \Bigl[ \Tr\Bigl( 
  \GW(t,\vec x)\, \GW^\dagger(t,\vec x\/) \Bigr)\Bigr]^2 \\
C(t) &=& \dsp \sum_{\vec x} \Tr\Bigl(
  \GW(t,\vec x\/) \, \GW^\dagger(t,\vec x\/)\,\,  
  \GW(t,\vec x\/) \, \GW^\dagger(t,\vec x\/)
\Bigr)
\ea
\eeq
where the trace is over color and spinor indices, and
$\GW(t,\vec x\/)$ is the quark propagator 
in a given gauge background from
a wall source at $t=0$ to the point $\vec x$ at time $t$. For smeared
correlators, we performed covariant smearing at the sink.
Note that the source for a pseudoscalar meson is $\bar\psi\ga_5\psi$, and the
inverse propagator $\GW^{-1}(t,\vec x\/) = \ga_5 \GW(t,\vec x\/) \ga_5$,
so no factors of $\ga_5$ appear in \eq{contractions}.

We construct the exotic and non-exotic correlators 
\beq\label{i02}
\ba{rcl}
J_\nonexotic(t) &=& \<D(t)\> + \half \<C(t)\> \\
J_\exotic(t) &=& \<D(t)\> - \<C(t)\>,
\ea
\eeq
where the angle brackets signify an average over the ensemble of
gauge field configurations.

\begin{figure}
\includegraphics[width=4in,angle=-90]{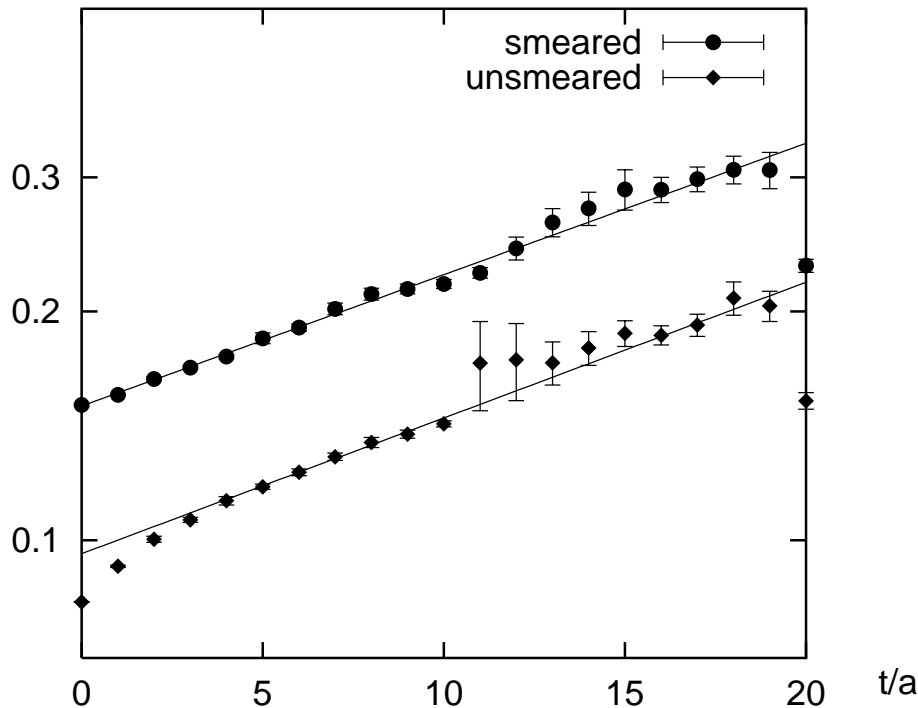}
\vspace{1mm}
\caption{
Ratio of wall-source correlators $R_\nonexotic(t)$ (see \eq{fitform}), 
on a log scale, for lattice spacing $a=0.4~\fm$, lattice size $L=2~\fm$.
The lattice was 40 spacings (10 fm) long, but only the first half is shown.
For the smeared data we fitted to an exponential using
$t=1$ to $17$. For the unsmeared we used $t=5$ to $17$.
The fitted values of $\dz$ agree.
}
\label{fig:plateau}
\end{figure}

To obtain the binding energy $\dz$ in the $I=0$ channel,
and the binding energy $\dt$ in the $I=2 $ channel, we
construct ratios of correlators and fit them to an exponential
\beq
\label{fitform}
\ba{c}
\dsp R_\nonexotic(t) = 
  \frac{J_\nonexotic(t)}{\<P(t)\>^2} \sim A \exp(-\dz t), \\[2ex]
\dsp R_\exotic(t) = 
  \frac{J_\exotic(t)}{\<P(t)\>^2} \sim B \exp(-\dt t).
\ea
\eeq
The ratios of correlators are expected to take the single exponential
form only at large $t$, after contributions from excited states have
died away. We followed the usual procedure of looking for a plateau,
and checking that smeared and unsmeared correlators give consistent
results.  The results for a typical case are shown in
Fig.~\ref{fig:plateau}.  There is no difficulty in identifying the
plateau and extracting $\dz$.

\section{Results and Discussion}
\label{section3}
\label{sec:results}
\subsection{Our Results}

\begin{figure}
\includegraphics[width=3.5in,angle=-90]{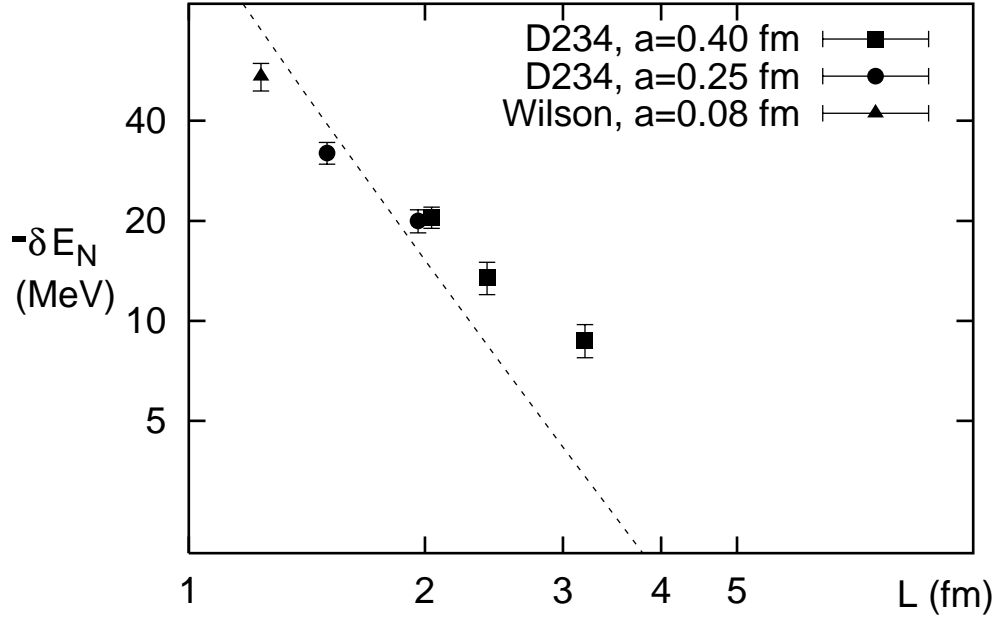}
\includegraphics[width=3.5in,angle=-90]{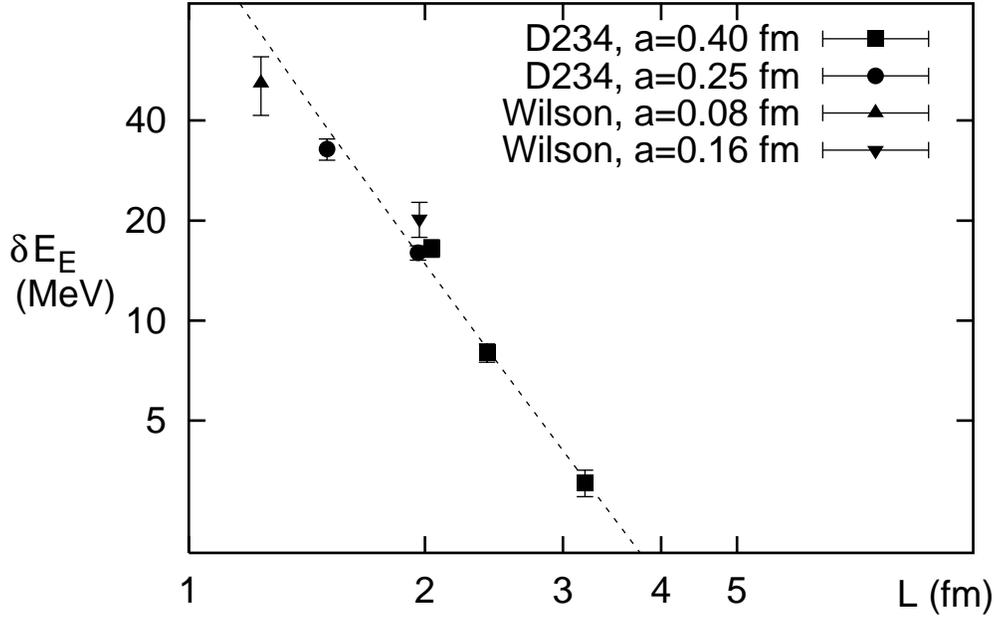}
\caption{ $P$-$P$ binding energy in non-exotic (N) and exotic 
(E) channels.
The data at $a=0.08~\fm$ are from \protect\cite{GPS2}; 
the data at $a=0.16~\fm$ are from \protect\cite{Fukugita}.
The $a=0.25~\fm$ and $a=0.4~\fm$ points at $L=2~\fm$
have been displaced slightly to either side in order to distinguish them.
The lines are fits to \protect\eq{luscher} (see table \ref{tab:fits}).
}
\label{fig:falloff}
\end{figure}

We measured $\delta E_{N}$ and $\delta E_{E}$ for several different
lattice spacings and sizes. Our results are shown in
Fig.~\ref{fig:falloff} along with previous results from
Refs.~\cite{GPS2,Fukugita}.  
The exotic and non-exotic channels appear
to scale differently as a function of $L$.
The exotic channel falls
like $1/L^3$, which is the expected form for a scattering
state, \eq{luscher}. A fit is given in Table \ref{tab:fits}, and
shown in the figure.  The non-exotic channel appears to depart from
$1/L^3$ falloff.  To be complete,
however, we have fitted the non-exotic data also to the form expected for
a scattering state.  The results are given in Table \ref{tab:fits} and
in the figure.

\begin{table}
\begin{tabular}{cccccc}
group & $\beta$ & $a^{-1}~(\MeV)$ & $\ka$ & $m_P/m_V$ & $m_P~(\MeV)$ \\
\hline
Gupta~\etal \cite{GPS2} & Wilson 6.0 
 & 2590(60)\cite{nrqcd} & 0.154 & ---  & 940(30) \\
Fukugita~\etal \cite{Fukugita} & Wilson 5.7 
& 1220(180)\cite{nrqcd}  & 0.164 & 0.740(8) & 620(90) \\
this work      & L\&W  1.719  
  & 790(10) \cite{land234}  & ---   & 0.756(5) & 840(11) \\
               & L\&W  1.157  
  & 495(4) \cite{land234}  & ---   & 0.756(4) & 790(6)
\end{tabular}
\vspace{2mm}
\caption{Lattice parameters for studies of $P$-$P$
scattering states. Lattice spacings are determined by
charmonium or upsilon $P\!-\!S$ splitting}
\label{tab:params}
\end{table}

The parameters of the lattice calculations are given in Table
\ref{tab:params}.  The lattice spacings were determined by quarkonium
$P\!-\!S$ splittings, using either charmonium or upsilon experimental
measurements to set the scale.

Although we only studied one quark mass, Gupta~\etal \cite{GPS2}
repeated their calculation for a lower quark mass, corresponding to
$m_P=770~\MeV$, and found that $\dz$ and $\dt$ were unchanged to
within statistical errors.  This suggests that studies of $\qqqq$
operators near $\qq$--$\qq$ thresholds are not too sensitive to quark
masses and leads us to combine the energy splittings from the
different calculations in Table~\ref{tab:params} on the same plot. For
the $\dt$ data we display both Gupta~\etal\  and Fukugita~\etal's
results with our own.  We enlarged the error bar on Fukugita~\etal's
point to 14\%, since Ref.~\cite{nrqcd} quoted a $14\%$ uncertainly in
measuring their lattice spacing, which arises from the discretization
errors involved in using an unimproved action on a coarse lattice. 
These are also apparent from the fact that Ref~\cite{Fukugita}'s
$m_{P}/m_{V}$ is close to ours, but its $m_P$ is significantly lower. 
For $\dz$ we do not use Fukugita~\etal's data, since they included the
annihilation diagrams, which we specifically exclude in order to see a
$\qqqq$ state.

Our results are consistent with those of Refs.~\cite{GPS2,Fukugita}, even
though we use much coarser lattices.  This supports our use of
Symanzik-improved glue and quark actions with tadpole-improved
coefficients.  As a further check on the validity of the improved
actions, we note that at $L=2~\fm$, where we performed a calculation at
two different lattice spacings for the same lattice volume, the
results for the two lattice spacings agree very well.  There is no
evidence of any discretization errors.

For the exotic $\qqqq$ system, the fit to \eq{luscher} is quite 
good, and the fitted scattering amplitude is remarkably similar to the result
expected in the chiral limit, $4f_P^2 T=1$.\footnote{Since we did not
calculate $f_P$ at our quark masses, we have used the value $f_P =
148$ MeV, derived from Ref.~\cite{GPS2}, Table 1.}  We conclude that 
there are no surprises in the exotic channel -- the interaction near 
threshold appears repulsive and the strength is close to that 
predicted by chiral perturbation theory.

The non-exotic $\qqqq$ system, however, does not fit the expected
scaling law at large $L$.  The fit to \eq{luscher} has a very large
$\chi^2$, and is so poor that the extracted amplitude $T$ is
meaningless.  Instead $\dz$ appears to be approaching a negative
constant at large $L$.  Instead of a scattering state, we appear to be
seeing a \emph{bound state} in the non-exotic channel.  Although our
data are suggestive, they are not conclusive.  It would be very
interesting to gather more data at $L\gtrsim 4$ fm, as well as at a
range of quark masses, in order to verify the existence of this new
state in the quenched hadron spectrum.

\begin{table}
\begin{tabular}{ccccc}
channel & amplitude $T(\MeV^{-2})$ & $\chi^2$/dof & dof&    $4 f_P^2 T$ \\
\hline
non-exotic & $1.20(5)\times 10^{-5}$ & 17.0 & 5 & --- \\ 
exotic     & $1.18(3)\times 10^{-5}$ & 3.0  & 6 & 1.03(3)
\end{tabular}
\vspace{2mm}
\caption{
Threshold scattering amplitudes $T$ obtained by
fitting lattice calculations of $\dz$ and $\dt$
to the scattering-state form \protect\eq{luscher}.
In the chiral limit, $4 f_\pi^2 T=1$.
The non-exotic channel data does not fit the scattering-state form, so
its fitted $T$ is meaningless.
}
\label{tab:fits}
\end{table}

\subsection{Interpretation and Discussion}

We have found evidence for a $\qqqq$ bound state just below threshold
in the non-exotic pseudoscalar-pseudoscalar $s$-wave.  In 2-flavor QCD
the bound state would correspond to an isosinglet meson coupling to
$\pi\pi$.  In 3-flavor QCD the non-exotic channel corresponds to an
entire nonet including two non-strange isosinglets and an isotriplet,
and two strange isodoublets (see Fig.~\ref{quark}b).  
We work with a large quark mass so our 
results are not directly applicable to $\pi\pi$ scattering, but they
do resemble physical $K\bar K$ scattering.\footnote{Although we work
in the $SU(3)_{\rm f}$ limit where all quark masses are equal.} The known
isosinglet $f_{0}(980)$ and isotriplet $a_{0}(980)$ mesons are obvious
candidates to identify with the non-exotic $\qqqq$ bound states we seem
to have found on the lattice.

We believe the quark mass dependence of the non-exotic $\qqqq$ state
is quite different from a standard $\qq$ lattice state.  In the
quenched approximation the masses of $\qq$ states like those shown in
Fig.~\ref{fig:qqbar} are roughly independent of $m_{P}$.  Note especially
that the masses are smooth as they cross the
threshold, $2m_{P}$.  In contrast, we believe that the $\qqqq$ state
we may have identified is strongly correlated with the $PP$ threshold
when the quark mass is large, and departs from it in a characteristic
way as the quark mass is reduced. 
(Indirect support for this comes from
Gupta~\etal's finding that their binding energy is independent of the
pseudoscalar mass.)
In particular, we believe that the
bound state will move off into the meson-meson continuum as $m_{P}$ is
reduced toward the physical pion mass.  

\begin{figure}
\includegraphics[width=3.5in,angle=-90]{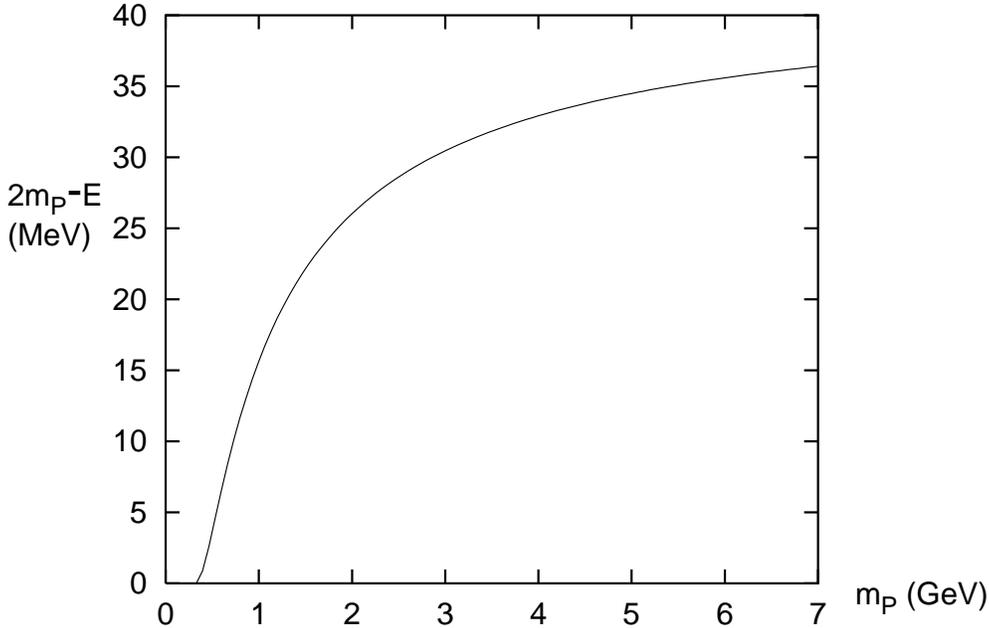}
\vspace{3mm}
\caption{Binding energy in MeV of the two-particle
state in our toy model \eq{toy}, as a function of particle mass in GeV.}
\label{toymodel}
\end{figure}

To explore the $m_{P}$ dependence of our results, we have made a toy
model based on a relativistic generalization of potential scattering. 
We write a Klein-Gordon equation for the $s$-wave relative meson-meson
wavefunction, $\phi(r)$,
\be\label{toy}
   -\phi^{\prime\prime}(r) + (2m_{P}-U(r))^{2}\phi(r)= E^{2}\phi(r),
   \label{kg}
\ee
with the boundary condition that $\phi(0)=0$.
For $U(r)=0$ the spectrum is a continuum beginning at $E=2m_{P}$ as
required.  In the non-relativistic limit $m_{P}\ll |U|$, 
\eq{kg} reduces to the
Schr\"odinger equation with an attractive potential $-U(r)$ (for
$U(r)>0$).  For sufficient depth and range, this potential will have a
bound state.  However,
as $m_{P}\to 0$, the potential term in \eq{kg} turns repulsive and
the bound state disappears.  Thus, if one keeps the depth and range of
$U$ fixed as one decreases $m_{P}$, the bound state moves out into the
continuum and disappears.  To be quantitative, we have taken a square
well, $U(r)=U_{0}$, for $r\le b$, and $U(r)=0$ for $r>b$.  We chose a
range $b=1/m_{\pi}\approx$ 1.4 fm, and adjusted $U_{0}$ such that 
the bound state has binding energy of 10 MeV when $m_{P}\sim800~\MeV$. 
The bound state does indeed move off into the continuum (first as a
virtual state) when $m_{P}$ goes below $330~\MeV$. The behavior of the
bound state in this toy model is shown in Fig.~\ref{toymodel}.  Note
this toy model is not meant to be definitive%
\footnote{
We could have chosen a different relativistic
generalization of the Schr\"odinger equation which would have 
preserved the bound state
as $m_{P}\to 0$.  For example, we could have replaced
$(2m_{P}-U)^{2}$ by $2m_{P}^{2}-2m_PU_1-U_2^2$, and fine-tuned
$U_1$ and $U_2$ to provide binding at arbitrarily low $m_P$.
}
but it illustrates the expected behavior of a $P$--$P$ bound state:
tracking $2m_P$ with roughly constant binding energy as $m_P$ falls, then
unbinding at some critical  $m_P$.

On the basis of our lattice computation and the $m_{P}$ dependence
suggested by our toy model, we believe it is possible that \emph{all}
the phenomena associated with the light scalar mesons are linked to
$\qqqq$ states.  The narrow $\0$ isosinglet $f_{0}(980)$ and
isotriplet $a_{0}(980)$ mesons near $K\bar K$ threshold can be
directly identified with $\qqqq$ lattice bound states (top line of
Fig.~\ref{quark}b).  The broad $\kappa(900)$ and $\sigma(600)$ (middle
and bottom lines of Fig.~\ref{quark}b) couple to low mass ($\pi\pi$ or
$\pi K$) channels.  We speculate that they are to be identified as the
continuum relics of the same objects which appear as bound states of
heavy quarks.

Of course, a thorough examination of this question would require
implementing flavor $SU(3)$ violation by giving the strange quark a
larger mass.  This would mix and split the isoscalars, shift the other
multiplets (see Fig.~\ref{quark}b), and dramatically alter
thresholds. For example, the $I=1$ $\qqqq$ state couples both to
$K\bar K$ and $\pi\eta$ (through the $\bar s s$ component of the
$\eta$) in the quenched approximation. The fact that the physical
$K\bar K$ and $\pi\eta$ thresholds are significantly different would
certainly affect the manifestation of bound states such as those
we have been discussing in the $SU(3)$-flavor-symmetric limit.

\subsection{Conclusions and Future Work}

We have presented evidence for previously unknown pseudoscalar meson
bound states in lattice QCD. Our results need confirmation. 
Calculations on larger lattices are needed, 
and variation with quark mass, lattice spacing, and discretization scheme
should be explored.

In the real world a $\0~\qqqq$ state may, depending on its flavor
quantum numbers, mix with $\0~\qq$ and glueball states.  
It seems natural to expect that for sufficiently
heavy quarks a bound state will remain, but
only full, unquenched lattice calculations can confirm this.

It is possible that an existing unquenched study of $\0~\qq$ operators may 
show some corroboration of our results.  In Ref.~\cite{UKQCD2}
the authors study $\qq$ sources with dynamical fermions.  Although 
their interest was in exploring the mixing of $\0~\qq$ with glueballs, 
there is nothing to stop their $\qq$ source from mixing with $\qqqq$.  
So they should be sensitive to the $\qqqq$ bound state we have 
identified.  It is therefore quite interesting that they report
a $\0$ state with an anomalously low mass $\sim 800~\MeV$.

If light $\qqqq$ states are, in fact, a universal phenomenon, and if
the $\sigma(600)$ is predominant\-ly a $\qqqq$ object, 
then the chiral transformation properties of
the $\sigma$ have to be re-examined.  The $\pi$ and
the $\sigma(600)$ are usually viewed as members of a (broken) chiral
multiplet.  In the naive $\qq$ model both $\pi$ and $\sigma$ are in
the $(\half,\half)\oplus(\half,\half)$
representation of $SU(2)_{L}\otimes SU(2)_{R}$ before symmetry
breaking.  In a $\qqqq$ model, as in the real world,
the chiral transformation properties of the $\sigma$ are not clear.

If the phenomena that we have discussed survive the introduction of
differing quark masses, then they will also have implications for
heavy quark physics. For example, there could be a $\0$ bound state
just below the decay threshold for two $D$ mesons in the charmonium spectrum.

Finally, we note that calculations similar to ours could be undertaken
in the meson-baryon sector and in other $J^{PC}$ meson channels.  It
has long been speculated that the $\Lambda(1405)$ is some sort of $KN$
bound state\cite{Pakvasa:1999zv} and $\qqqq$ states have been
postulated in other meson-meson partial waves.

\section{Acknowledgments}

We would like to thank the members of the CTP Phenomenology Club for 
the stimulating environment in which this project was conceived,
and Craig McNeile, Weonjong Lee, and Seyong Kim for discussions
of their published work.

\vspace*{1cm}

This work is supported in part by funds provided by the U.S.
Department of Energy (D.O.E.) under cooperative research agreement
\#DF-FC02-94ER40818.  The lattice QCD calculations were performed on
the SP-2 at the Cornell Theory Center, which receives funding from
Cornell University, New York State, federal agencies, and corporate
partners. The code was written by P. Lepage, T. Klassen, and M. Alford.


\end{document}